\documentclass[twocolumn,prl,showpacs,preprintnumbers,amsmath,amssymb]{revtex4}

\usepackage{graphicx}
\usepackage[usenames]{color}
\begin{document}

\title{Two-photon path-entangled states in multi-mode waveguides}

\author{Eilon Poem}
\email{eilon.poem@weizmann.ac.il}
\affiliation{Department of Physics of Complex Systems, Weizmann Institute of Science, Rehovot 76100, Israel}
\author{Yehonatan Gilead}
\affiliation{Department of Physics of Complex Systems, Weizmann Institute of Science, Rehovot 76100, Israel}
\author{Yaron Silberberg}
\affiliation{Department of Physics of Complex Systems, Weizmann Institute of Science, Rehovot 76100, Israel}

\date{\today}

\begin{abstract}
We experimentally show that two-photon path-entangled states can be coherently manipulated by
multi-mode interference in multi-mode waveguides. By measuring the output two-photon spatial correlation function versus the phase of the input state, we show that multi-mode waveguides perform as nearly-ideal multi-port beam splitters at the quantum level, creating a large variety of entangled and separable multi-path two-photon states.
\end{abstract}

\pacs{42.50.Dv, 03.67.Bg, 42.79.Gn}

\maketitle

Quantum states of photons distributed between several optical paths
play the central role in linear 
optical quantum computation~\cite{KLM01}. In particular, entangled states, where no
decomposition into a product of either single-path or single-photon
states exists, are of great importance due to their non-local
nature~\cite{Bennet93,Jennewein00}. Entangled photons states 
are generated either by non-linear processes in 
crystals~\cite{Kwiat95}, by cascaded emission of single quantum
emitters~\cite{Akopian06}, or by quantum interference of
non-classical light on a beam splitter~\cite{Fattal04,Afek10_1}.

Quantum interference in arrays of beam-splitters (BSs) is used in
many linear optical quantum computation schemes to create and
manipulate multi-photon multi-path states~\cite{ZZH97,Sagi03,Lim05}.
The simplest example is the Hong-Ou-Mandel (HOM)
interference~\cite{HOM87}. In this fundamental effect, two identical photons
input each on a different port of a 50:50 BS, cannot exit in two
different ports due to destructive interference of the two possible
paths. Therefore, they always exit bunched together on either one of
the output ports, in an entangled state.
A major obstacle on the way
to the implementation of quantum optical circuits is the large
number of BSs required, and the increasing complication of their
alignment. One way to overcome this problem is through miniaturization of the optical circuites. Indeed, quantum interference has been recently demonstrated in integrated optical circuites composed of evanescently coupled single-mode waveguides embedded in solids~\cite{Peruzzo10,Sansoni12,Obrien_arXiv}. A conceptually different route towards robust implementation of quantum optical circuites may come in the form
of \emph{multi-mode} interference (MMI) devices~\cite{Soldano95}. These
compact replacements for BS arrays, usually based on planar
multi-mode waveguides (MMWs), are already used extensively in modern
classical optical communication networks. They have also been proposed to be useful for creation and detection of multi-photon states~\cite{Lim05}, as they naturally implement Bell multiport BSs~\cite{ZZH97}. A step towards their use
in quantum networks was recently made with the demonstration of HOM
interference of two separated photons in an MMW~\cite{Peruzzo11}.
The question still remains, however, whether entangled states, which unlike separated photons, carry relative phase information, can be coherently manipulated by MMWs.

Here we experimentally show that the answer to this question is positive. We utilize MMI in a two-mirror, tunable planar MMW~\cite{Poem11} to implement multi-port
BS arrays of up to 5 input and 5 output ports, and explore the
propagation of non-classical, path-entangled two-photon states through
them. We measure the two-photon correlations at the output, and find that, for any relative phase between the two paths, they agree very well with the multi-path, two-photon states expected at the output of ideal multi-port BSs.

\begin{figure}[tbh]
\centering
\includegraphics[width=0.48\textwidth]{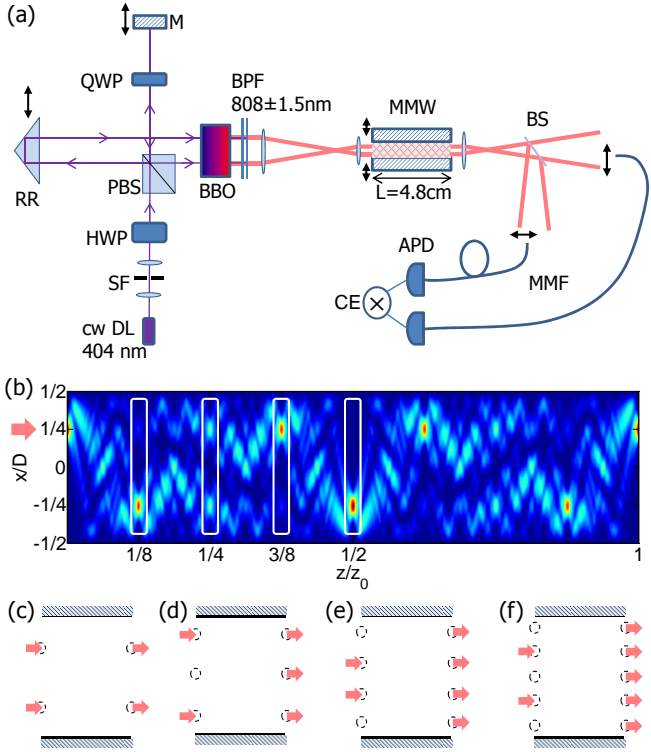}
\caption{(a) The experimental setup. DL, 404 nm cw Diode laser; SF, Spatial filter; HWP,
Half wave plate; PBS, Polarizing beam splitter; RR, Retroreflector; QWP, Quarter
wave plate; M, Mirror; BBO, $\beta$-Barium-Borate crystal; BPF, Band-pass spectral filter;
MMW, Multimode waveguide; BS, beam splitter; MMF, Multimode fiber; APD, Avalanche
photodiode; CE, Correlation electronics. The thin purple (wide red) lines represent
the pump (down conversion) beams. Movable parts are marked by black double
arrows. (b) Calculated intensity distribution in an MMW for an input beam centered about $x=D/4$. The white boxes mark propagation lengths where various 2$\times$2 BSs are implemented. (c)-(f) Schematic diagrams showing the configurations of input and output ports used in the experiments.}\label{fig:1}
\end{figure}

The experimental system is described in Fig.~\ref{fig:1}(a). States of
the form
$\psi_2^{\phi}=\frac{1}{\sqrt{2}}\left(|2,0\rangle+e^{i\phi}|0,2\rangle\right)$
are created by first splitting a 404 nm continuous-wave (cw) diode-laser beam into two
arms. The relative phase between the beams, $\phi$, is controlled and stabilized by a piezoelectrically 
movable mirror on one arm [see Fig.~\ref{fig:1}(a)]. The state in the two arms at this point is $|\alpha,e^{i\phi}\alpha\rangle$, where $\alpha$ represents a coherent state of average photon number $|\alpha|^2$. Note that the single photon
part of this state is
$\psi_1^{\phi}=\frac{1}{\sqrt{2}}\left(|1,0\rangle+e^{i\phi}|0,1\rangle\right)$.
The two beams then undergo type I collinear degenerate spontaneous parametric
down-conversion (SPDC) in a properly oriented  2~mm long $\beta$-Barium-Borate (BBO)
crystal. In this process, each 404~nm photon transforms into two 808~nm photons propagating together. Since SPDC is much faster than the optical period, the relative phase between the two paths is kept as it was before the BBO crystal. Spectral filters are used to eliminate
all remaining photons outside a 3~nm band 
centered at 808~nm. The two photon part of the state after the BBO and the spectral filters is thus
$\psi_2^{\phi}$. There is no single-photon part, and for low enough pump powers, the content of the next order - the four photon part, is negligible.

The beams are then inserted into an MMW made of two parallel metallic
mirrors~\cite{Poem11}.
The distribution of one component of the optical field between the mirrors, assuming perfect reflection and paraxial propagation, is given by~\cite{Soldano95},
\begin{equation}\label{Eq:field}
    E(x,z)=e^{-ikz}\sum_{n=1}^{\infty}{A_n\sin\left[n\pi\left(x-D/2\right)/D\right]e^{i2\pi n^2z/z_0}},
\end{equation}
with $k=2\pi/\lambda$ and $z_0=8D^2/\lambda$,
where $D$ is the width of the MMW, $\lambda$ is the wavelength of the incident light, and the constants $A_n$ are determined by the incident field distribution. 
From Eq.~\ref{Eq:field} it can be shown that for any input, full imaging occurs at \mbox{$z=z_0$}, reflection about the center of the MMW occurs at \mbox{$z=z_0/2$}, and two-way, equal beam splitting with a relative phase of $\pi/2$ ($-\pi/2$) occurs at \mbox{$z=z_0/4$} (\mbox{$z=3z_0/4$}).  These effects are known as \emph{general} interference~\cite{Soldano95}. Unequal BSs, and BSs of more than two ports are also possible, but only for specific positions of a localized input beam. This is known as \emph{restricted} interference~\cite{Soldano95}. For example, for an input beam localized about \mbox{$x=\pm D/4$}, an unequal two-way BS will be realized at \mbox{$z=(2m-1)z_0/8$}, where $m$ is an integer. This is readily seen in Fig.~\ref{fig:1}(b), where the intensity distribution inside the MMW, calculated for a Gaussian input beam localized about \mbox{$x=D/4$}, is presented. Generally, an $N$$\times$$N$ BS would be generated by restricted interference for beams localized about \mbox{$x=(2p-1-N)D/2N$}, for \mbox{$z=qz_0/4N$}, where $p$ and $q$ are integers, and $p$ goes from 1 to $N$. For an even $q$, the BS is equal. Using Eq.~\ref{Eq:field}, the transition matrix, \mbox{$\mathbf{T}^{(1)}_{N,q}=\left(\mathbf{T}^{(1)}_{N,1}\right)^q$}, can be calculated for any
$N$~\cite{Heaton99}.
This matrix applies for coherent beams as well as for a single photon. In order to obtain the transition matrices for states of $M$ identical photons, one should sum up all the products of the single-photon matrix elements relevant to the transition between an initial, $\vec{\nu}$, and a final, $\vec{\mu}$, $M$-photon configuration,
\begin{equation}\label{Eq:BS_matrix_M}
    \left(\mathbf{T}^{(M)}_{N,q}\right)_{\vec{\mu},\vec{\nu}}=\sqrt{\frac{N_{\vec{\mu}}}{N_{\vec{\nu}}}}\sum_{\{\vec{\sigma}_{\vec{\nu}}\}}{\prod_{j=1}^M{\left(\mathbf{T}^{(1)}_{N,q}\right)_{\mu_j,\sigma_{\vec{\nu}j}}}},
\end{equation}
where $\vec{\sigma}_{\vec{\nu}}$ represent permutations of the initial configuration $\vec{\nu}$, and $N_{\vec{\nu}}$ ($N_{\vec{\mu}}$) is the number of all different permutations of the initial (final) configuration.
The calculation of these $N$-port, $M$-photon transition matrix elements is, in general, a computationally difficult task, equivalent to the calculation of the permanent of a matrix. However, it is feasible for small enough $M$ and $N$, and for \mbox{$N=2$} there even exists an analytic expression for any $M$~\cite{Campos89}. Given an initial wavefunction (in terms of the initial amplitudes for each distribution of the $M$ photons among the $N$ input ports), the final state, $\psi_{out}$, is readily found, and from it, any correlation function can be calculated. In particular, for a general two-photon state, the two-photon correlation probability between ports $m$ and $n$ is just
\begin{equation}\label{Eq:Corr_2}
    P^{(2)}_{m,n}=|\langle m,n|\psi_{out}\rangle|^2,
\end{equation}
where
$|m,n\rangle$ is a state where one photon is on the $m^{th}$ output port, and the other is on the $n^{th}$ output port.

In our experimental setup, the length of the MMW is fixed by the length of the mirrors, $L$. However, the adjustment of the \emph{relative} length,
\mbox{$\zeta=L/z_0=L\lambda/8D^2$}, is possible through the adjustment of the separation between the mirrors. For the experiments presented below the MMW was adjusted to function as 2$\times$2, 3$\times$3, 4$\times$4, and 5$\times$5 BSs. Figs.~\ref{fig:1}(c)-(f) show the corresponding configurations of input and output ports used in these experiments.
We note that due to the finite conduction of the mirrors the MMW has a slightly different relative length for different polarizations. To avoid any possible complications that may arise, the polarization of the incident photons (in both beams) was set parallel to the mirrors, such that only TE modes were excited.
In order to measure two-photon correlations at the output of the MMW, the light at its output facet is split and imaged on two
multi-mode fibers, each connected to a single photon detector (Si
avalanche photodiode). 
The digital signals from the two detectors are input into
correlation electronics that shorten and multiply them, yielding the rate of coincidence events in a 7~ns time window~\cite{WhitmanCCU}. As the fibers can be placed in front of any pair
of output beams [see Fig.~\ref{fig:1}(a)], both auto- and
cross-correlations can be measured. Since in this method, the beams are split into two also when cross-correlations are measured, the measured rate of cross-correlations would be half that predicted using Eq.~\ref{Eq:Corr_2}. For the comparison of theory and experiment we therefore define the modified correlation probability,
\begin{equation}\label{Eq:Mod_Corr_2}
    C^{(2)}_{m,n}=P^{(2)}_{m,n}/(2-\delta_{m,n}).
\end{equation}

\begin{figure}[htbp]
  \centering
  \includegraphics[width=0.48\textwidth]{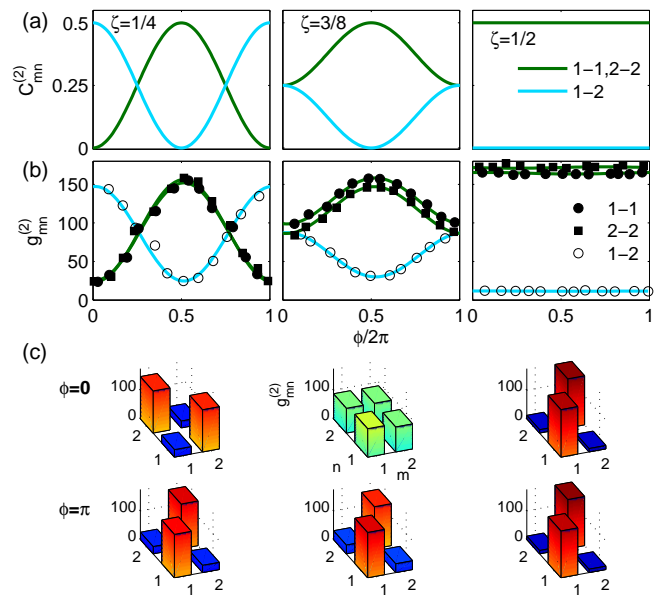}
\caption{(a) Two-photon correlations vs. the phase, $\phi$, calculated using Eq.~\ref{Eq:Mod_Corr_2} for relative waveguide lengths of 1/4 (left), 3/8 (center), and 1/2
(right), where the MMW functions as different types of 2$\times$2
BSs, as illustrated in Fig.~\ref{fig:1}(c). The correlation between ports m and n is marked by `m-n'. (b) Measured correlation functions. The symbols are the measured values and the lines are best
fitted 2$\pi$ period sinusoidal functions. (c) Corresponding
two-photon correlation maps extracted from the fitted curves for $\phi=0$ (top) and $\phi=\pi$
(bottom).}\label{fig:2}
\end{figure}

Fig.~\ref{fig:2}(a) presents the calculated modified two-photon correlation
probabilities (Eq.~\ref{Eq:Mod_Corr_2}) versus the phase $\phi$, for three
different relative propagation lengths at which 2$\times$2 BSs are implemented [see Fig.~\ref{fig:1}(b) and (c)]: $\zeta=1/4$ - a 50:50 BS with a relative phase of $\pi/2$, $\zeta=3/8$ - a $\cos^2\pi/8:\sin^2\pi/8$ ($\sim$85:15) BS with a relative phase of $\pi/2$, and $\zeta=1/2$ - a reflection about the center of the MMW (a `0:100 BS'). Note that all three cases are a part of the same series of transformations given by $\left(\mathbf{T}^{(1)}_{2,1}\right)^q$, where $\mathbf{T}^{(1)}_{2,1}$ is a 15:85 BS implemented for $\zeta=1/8$, and that for $q=8$ ($\zeta=1$), full imaging is obtained~\cite{Heaton99,Poem11}. The state $\psi_2^0$ is invariant only under reflection, and thus returns to
its initial form only when full reflection occurs - once every four applications of $\mathbf{T}^{(1)}_{2,1}$ (e.g. at $\zeta$=1/2). Note that at the middle of this period
(on the 50:50 BS at $\zeta$=1/4) this state transforms into two separated photons,
exhibiting inverse HOM interference. In contrast, the state
$\psi_2^{\pi}$ is left unchanged (up to a global phase) for each and
every application of $\mathbf{T}^{(1)}_{2,1}$. This is because this state is invariant under \emph{any} unitary two-way beam
splitter with a relative phase of $\pm\pi/2$~\cite{Yurke86}.

\begin{figure}[tb]
  \centering
  \includegraphics[width=0.48\textwidth]{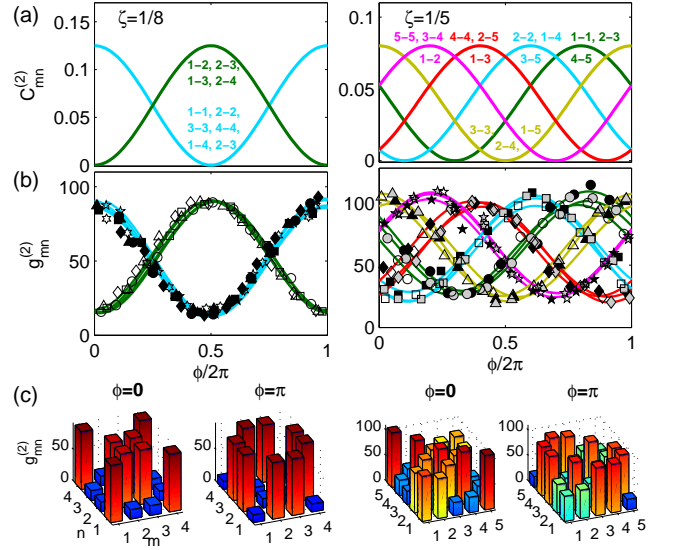}
\caption{(a) Calculated two-photon correlations vs. the phase, $\phi$, for an
equal 4$\times$4 multi-port BS (left), implemented by setting the
relative length of the MMW to 1/8, and an equal 5$\times$5
multi-port BS (right), for which the relative length is 1/5. The locations of the input and output ports are illustrated in Fig.~\ref{fig:1}(e) and (f), respectively. The output ports configurations (`m-n') of all correlations described by a certain curve are listed next to it. (b) Measured correlation functions. Full
black symbols represent auto correlations, while gray-filled and
empty symbols represent cross-correlations. The solid lines are sinusoidal fits. (c) Corresponding
two-photon correlation maps for $\phi=0$ and
$\phi=\pi$.}\label{fig:4}
\end{figure}

Fig.~\ref{fig:2}(b) shows the measured correlation functions, $g^{(2)}_{mn}$ (symbols), defined as the
rate of coincidence events between output ports $m$ and $n$, normalized by the expected accidental
coincidence rate, $r_a=r_mr_nw$, $r_m$ being the
single-count rate in port $m$, and $w=7$ ns the coincidence measurement time
window. The total integration time is 7 sec
for each data point. The lines are the best fitted $2\pi$-period sinusoidal functions. The measured background after the MMW (right panel of Fig.~\ref{fig:2}(b), empty circles) is $\sim$10 times larger than the expected accidental coincidence rate. By blocking either one of the input beams, we have verified that indeed $\sim$90\% of this background is due to a small overlap between the collecting fiber and the neighboring output port. This is also confirmed by the increase of the background level when the beams are brought closer (see, e.g., Fig.~\ref{fig:4}(b) below). The bare visibility of the oscillations measured for the 50:50 BS configuration (Fig.~\ref{fig:2}(b), left panel) is 72$\pm$3\%, violating the classical bound of 50\%~\cite{Afek10_2} by more than 7 standard deviations. This indicates that the light after the MMW is still entangled, even without background reduction. With the reduction of the measured background, the visibility reaches 83$\pm$3\%.

Fig.~\ref{fig:2}(c) presents `correlation
maps' for each of the relative lengths, showing the elements $g^{(2)}_{mn}$ for two initial phases, $\phi=0$ (top) and $\phi=\pi$ (bottom), as extracted from the fits to the measured data. For each phase, the
expected periodicity with respect to the relative length of the MMW
is clearly seen.
\begin{figure}[htbp]
  \centering
  \includegraphics[width=0.48\textwidth]{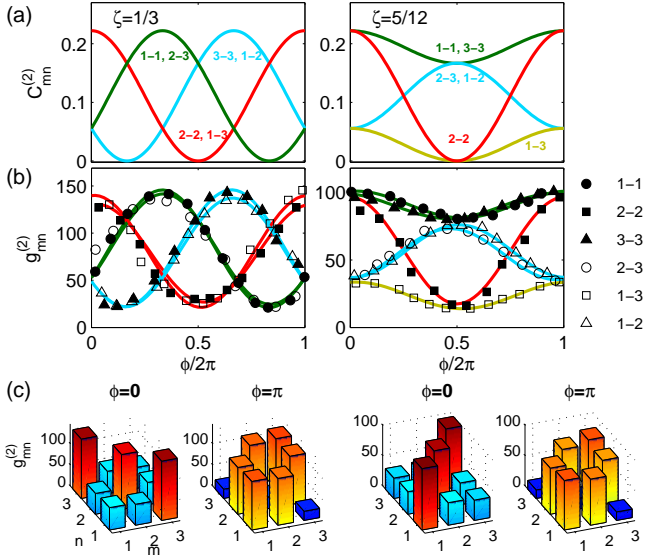}
\caption{(a) Calculated two-photon correlations vs. the phase, $\phi$, for two
cases of 3$\times$3 multi-port BSs implemented by setting the
relative length of the MMW to 1/3 (left), and 5/12 (right). The locations of the input and output ports are illustrated in Fig.~\ref{fig:1}(d). The port configurations (`m-n') of all correlations described by a certain curve are listed next to it. (b) The measured correlations (symbols) and their sinusoidal fits (lines). (c) Correlation maps for $\phi=0$ and
$\phi=\pi$.}\label{fig:3}
\end{figure}

We proceed to examine 3$\times$3 BSs. The input and output ports used are as illustrated in Fig.~\ref{fig:1}(d). Fig.~\ref{fig:3} shows the
two possible cases. On the left we present the calculated (a) and measured (b) correlations
of a 3$\times$3 equal BS ($\zeta=1/3$), while on the right the
correlations for an unequal BS ($\zeta=5/12$) are
presented.
For the equal BS, the six possible two-port correlation measurements divide into three pairs,
oscillating with a phase difference of $2\pi/3$ between them. Each
pair contains an autocorrelation of one port and the cross
correlation of the other two ports. There are six special phases
where two such correlations pairs meet. For three of them there is
an enhancement of one pair over the other two, while for the other
three, one pair is completely suppressed.  This is further
visualized in Fig.~\ref{fig:3}(c), where the correlation maps for
$\phi=0$ and $\phi=\pi$ are shown.
In contrast to the equal 3$\times$3 BS, the phase dependencies of the correlations in the unequal
3$\times$3 BS show only two different phases of oscillation and only partial visibilities.
However, the two cases can be related one to the other if one notes
that the unequal BS is composed of an equal 2$\times$2 BS
($\zeta=1/4$) on the two outer ports, followed by an equal
3$\times$3 BS of $\zeta=1/6$, different from that of $\zeta=1/3$
only in the order of the relative phases~\cite{Heaton99}. This is
best seen in the correlation maps shown in Fig.~\ref{fig:3}(c). For
$\phi=0$, the cross-correlation between the two outer arms switch
values with the corresponding auto-correlations, while for
$\phi=\pi$ there is no difference between the two 3$\times$3
BSs, due to the invariance of $\psi_2^{\pi}$ on 2$\times$2 $\pi/2$ BSs.

To examine the effect of even more complex BSs on the states
$\psi_2^{\phi}$, we performed measurements for equal 4$\times$4
and 5$\times$5 BSs. The input and output ports used are as illustrated in Figs.~\ref{fig:1}(e) and (f), respectively. Fig.~\ref{fig:4}(a) [(b)] presents the calculated [measured] phase
dependence of these correlations, and Fig.~\ref{fig:4}(c) shows the
corresponding correlation maps for $\phi=0$ and $\phi=\pi$.
In the case of the 4$\times$4 equal BS, all the auto correlations
and the cross-correlations of symmetric ports oscillate together
with a phase of 0, and all the cross-correlations of asymmetric
ports oscillate together with a phase of $\pi$. The state created
for $\phi=0$ therefore contains no asymmetric cross correlations,
while that created for $\phi=\pi$ contains only asymmetric cross
correlations. This is visualized in the correlation maps of
Fig.~\ref{fig:4}(c).
For the equal 5$\times$5 BS, analogously to the equal
3$\times$3 BS, the 15 correlations group in 5 triplets that
oscillate with phase differences of $2\pi/5$. Here too, some
correlations can be either relatively enhanced or completely
suppressed, as visualized in Fig.~\ref{fig:4}(c).

In summary, we experimentally show, for the first time, that path-entangled
quantum light can be coherently manipulated by multi-port BSs, naturally implemented by MMI in MMWs. The manipulation is shown to maintain the information on the relative phase between the two paths. This persists even for complex manipulations with a high number of output ports, allowing for the robust creation of a large space of two-photon multi-path states, controlled by the initial phase. MMI is thus shown to be a robust and simple approach for the implementation of various quantum optical circuites of high complexity. 

We thank Yaron Bromberg, Yoav Lahini, and Yonatan Israel for their help. The financial support of the Minerva
Foundation, the European Research Council, and the Crown Photonics Center is
gratefully acknowledged.


\end{document}